\documentclass[prb,superscriptaddress,amsmath,amssymb,floatfix,margin=1in,super,longbibliography,twocolumn]{revtex4-1} 

\usepackage[dvips]{graphicx}
\usepackage{xspace}
\usepackage[usenames,dvipsnames]{xcolor}
\usepackage{array}
\usepackage{ulem}
\usepackage{hyperref}
\usepackage[english]{babel}
\usepackage{csquotes}
\usepackage{multirow}
\usepackage{dblfloatfix}
\usepackage{amsmath}

\setcitestyle{super}

\begin{document}

\title{Nonequilibrium quasistationary spin disordered state in the Kitaev-Heisenberg magnet $\alpha$-RuCl$_3$}

\newcommand{\phii}{Institute of Physics II, University of Cologne, D-50937 Cologne, Germany}
\newcommand{\mineralogy}{Institute of Geology and Mineralogy, University of Cologne, Z\"{u}lpicher Stra{\ss}e 49b, D-50674 Cologne, Germany}
\newcommand{\thp}{Institute for Theoretical Physics, University of Cologne, D-50937 Cologne, Germany}
\newcommand{\augsburg}{Experimental Physics V, Center for Electronic Correlations and Magnetism, University of Augsburg, 86159 Augsburg, Germany}
\newcommand{\chisinau}{Institute of Applied Physics, MD 2028, Chisinau, Republic of Moldova}

\newcommand{\changer}[1]{\textcolor{black}{#1}}
\newcommand{\xx}{\mathbf{x}}
\newcommand{\yy}{\mathbf{y}}
\newcommand{\rr}{\mathbf{r}}
\newcommand{\res}{\widetilde{\phi}}
\newcommand{\dd}{\mathrm{d}}
\newcommand{\ee}{\mathrm{e}}
\newcommand{\kk}{\mathbf{k}}
\newcommand{\qq}{\mathbf{q}}
\newcommand{\vecpi}{\boldsymbol\Pi}
\newcommand{\vecphi}{\boldsymbol\phi}
\newcommand{\vecvarphi}{\boldsymbol\varphi}
\newcommand{\sgn}{{\rm{sgn}}}

\newcommand{\ale}[1]{{\bf \textcolor{blue}{#1}}}

\newcommand{\RC}{RuCl$_{3}$\@\xspace}
\newcommand{\aRC}{$\alpha$-RuCl$_{3}$\@\xspace}

\newcommand\redsout{\bgroup\markoverwith{\textcolor{black}{\rule[0.5ex]{2pt}{0.4pt}}}\ULon}

\author{R.~B.~Versteeg}
\email[Corresponding author: ]{rolf.versteeg@epfl.ch}
\altaffiliation[Current address: ]{Laboratoire de Spectroscopie Ultrarapide and Lausanne Centre for Ultrafast Science (LACUS), ISIC-FSB, \'Ecole
Polytechnique F\'ed\'erale de Lausanne, CH-1015 Lausanne, Switzerland. }
\affiliation{\phii} 
\author{A.~Chiocchetta}
\affiliation{\thp}
\author{F.~Sekiguchi}
\affiliation{\phii} 
\author{A.~I.~R.~Aldea}
\affiliation{\phii} 
\author{A.~Sahasrabudhe}
\affiliation{\phii} 
\author{K.~Budzinauskas}
\affiliation{\phii} 
\author{Zhe~Wang}
\affiliation{\phii} 
\author{V.~Tsurkan}
\affiliation{\chisinau}
\affiliation{\augsburg}
\author{A.~Loidl}
\affiliation{\augsburg}
\author{D.~I.~Khomskii}
\affiliation{\phii} 
\author{S.~Diehl}
\affiliation{\thp}
\author{P.~H.~M. van Loosdrecht}
\email[Corresponding author: ]{pvl@ph2.uni-koeln.de}
\affiliation{\phii}

\date{\today}

\begin{abstract}
Excitation by light pulses enables the manipulation of phases of quantum condensed matter. Here, we photoexcite high-energy holon-doublon pairs as a way to alter the magnetic free energy landscape of the Kitaev-Heisenberg magnet \aRC, with the aim to dynamically stabilize a proximate spin liquid phase. The holon-doublon pair recombination through multimagnon emission is tracked through the time-evolution of the magnetic linear dichroism originating from the competing zigzag spin ordered ground state. A small holon-doublon density suffices to reach a spin disordered state. The phase transition is described within a dynamic Ginzburg-Landau framework, corroborating the quasistationary nature of the transient spin disordered phase. Our work provides insight into the coupling between the electronic and magnetic degrees of freedom in \aRC and suggests  a  new  route  to  reach a proximate spin liquid phase in Kitaev-Heisenberg magnets.
\end{abstract}

\maketitle

\section*{Introduction}

Light can be utilized as a tool to manipulate and engineer novel phases in quantum materials~\cite{basov2017towards}. In particular, excitation via intense light pulses has been used to create nonequilibrium states of matter nonexistent at thermal equilibrium, such as transient superconductivity in underdoped cuprates~\cite{fausti2011light}, metastable ferroelectricity in SrTiO$_3$~\cite{Nova2019,Li2019}, and unconventional charge-density wave order in LaTe$_3$ \cite{Kogar2020}. The light pulses excite a transient population of quasiparticles or collective excitations, which acts as a dynamic parameter
to alter the material's free-energy landscape. 
For sufficiently strong excitation densities, a nonequilibrium phase transition can eventually occur~\cite{Kogar2020,teitelbaum2019,dolgirev2019universal}.
By the same token, intense pulsed light holds
promise to manipulate the spin state of frustrated magnets~\cite{balents2010spin,knolle2019field}. These materials, in fact, can host exotic and elusive phases, such as spin liquids (SL). Whereas SLs harbors rich many-body phenomena resulting from spin frustration and possible spin fractionalization~\cite{kitaev2006anyons,broholm2020quantum}, these phases often compete with a magnetically ordered ground state, which is typically energetically favoured.
Pulsed light excitation can then provide a mechanism to tip the energetic balance away from the magnetically ordered ground state towards a nonequilibrium proximate spin liquid phase.

We explore this concept for the Kitaev-Heisenberg frustrated magnet, a type of Mott insulator with a layered honeycomb structure and strong spin-orbit coupling~\cite{jackeli2009,chaloupka2013,takagi2019concept}. For these materials the large spin-orbit interaction leads to a sizeable bond directional spin exchange, whereas the symmetric Heisenberg exchange cancels out by virtue of the edge-sharing octahedra geometry, making them promising candidates for Kitaev physics~\cite{jackeli2009,chaloupka2013,takagi2019concept}. Still, the remaining Heisenberg interaction, present due to small structural distortions away from the ideal honeycomb structure, \cite{johnson2015} is an adversary to spin liquid formation, and generally favors a spin-ordered ground state~\cite{chaloupka2013,alpichshev2015,nembrini2016}. %
By modulating spin entropy through finite temperature effects ~\cite{do2017majorana,sandilands2015raman} or by adding external magnetic fields ~\cite{kasahara2018majorana}, one can however stabilize proximate or field-induced 
spin liquid phases at thermal equilibrium.
These spin liquid realizations show emergent behavior expected for the pure Kitaev spin liquid,\cite{kitaev2006anyons} most notably, fractionalized particle statistics \cite{sandilands2015raman} and quantized conduction phenomena.\cite{kasahara2018majorana}

A case in point is \aRC. This honeycomb Mott insulator has nearly-ideal $j_{\rm eff}$\,$=$\,$\tfrac{1}{2}$ isospins in highly symmetric octahedra,~\cite{plumb2014,agrestini2017}  making it possibly the most promising Kitaev spin liquid host studied to date~\cite{do2017majorana,sandilands2015raman,kasahara2018majorana}.
The ($B$,$T$)-plane in Fig.\,\ref{fig:phasediagram} provides the \textit{equilibrium} phase diagram as a function of magnetic field and temperature. Below $T_{\rm N}$\,$\approx$\,$7$K, the isospins couple in a zigzag fashion, consistent with the types of magnetic order captured by the Kitaev-Heisenberg model.\cite{chaloupka2013}
Strong short-range spin correlations persist between $T_{\rm N}$ and the crossover temperature $T_{\rm H}$\,$\approx$\,$100$\,K, hinting at the formation of a proximate spin liquid (pSL) phase within this intermediate temperature regime.
\cite{do2017majorana,banerjee2016proximate,winter2018}
Above $100$\,K thermal fluctuations bring the system into a conventional paramagnetic phase.
An additional tuning parameter is provided by an in-plane magnetic field. A field of $B_{\rm c}$\,$\approx$\,$7$\,T is sufficient to destabilize the zigzag order.
For fields between $7$\,-\,$8$\,T a much-debated field-induced SL is then stabilized, \cite{kasahara2018majorana} whereas for higher fields a quantum disordered state with partial field alignment of the effective moments forms.\cite{johnson2015,sears2017phase,sahasrabudhe2019high} 

\begin{figure}[h!]
\center
\includegraphics[width=2.75in]{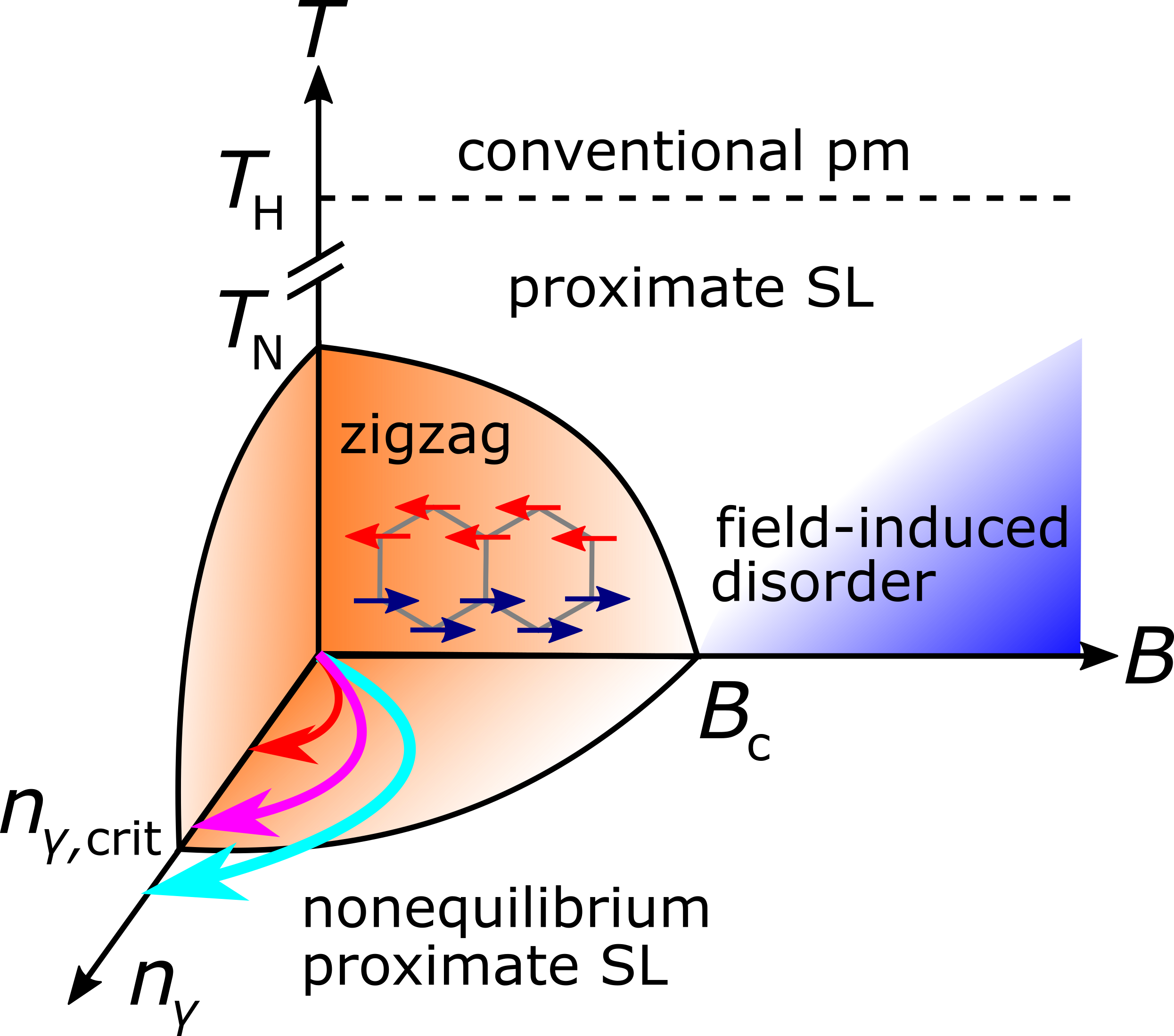} 
\caption{A nonequilibrium dimension to \aRC's magnetic phase diagram. 
The ($B$,$T$)-plane sketches the \textit{equilibrium} magnetic phase diagram. Photoexcited holon-doublon pairs $n_{\gamma}$ form a new \textit{nonequilibrium} parameter.
For small (red) to intermediate (magenta) quenches the system stays inside the zigzag ordered phase. Above a critical density $n_{\gamma,\rm crit}$ a nonequilibrium proximate spin liquid state may be induced (light blue arrow).}
\label{fig:phasediagram}
\end{figure}

In this work, we report on the observation of a transient long-lived spin disordered state in the Kitaev-Heisenberg magnet \aRC induced by pulsed light excitation. Holon-doublon pairs are created by photoexcitation above the Mott gap, and provide a new \textit{nonequilibrium} dimension to \aRC's magnetic free energy landscape and resulting phase diagram, as illustrated in Fig.\,\ref{fig:phasediagram}. The subsequent holon-doublon pair recombination through multimagnon emission leads to a decrease of the zigzag magnetic order. This is tracked through the magnetic linear dichroism (MLD) response of the system. For a sufficiently large holon-doublon density the MLD rotation vanishes, implying that the zigzag ground state is fully suppressed and that a long-lived transient spin-disordered phase is induced. The disordering dynamics of the zigzag order parameter is captured by a time-dependent Ginzburg-Landau model, corroborating the nonequilibrium quasistationary nature of the transient phase. Our work provides insight into the coupling between high-energy electronic and low-energy magnetic degrees of freedom in \aRC and suggests a new route to reach a proximate spin-liquid phase in honeycomb Mott insulators with residual interactions beyond the bond-directional Kitaev exchange.

\section*{Results and discussion}

The photoinduced change in reflected polarization rotation from \aRC was measured as a function of temperature and photoexcitation density. The sample is excited above the $\Delta_{\rm MH}$\,$\sim$\,$1.0$\,eV Mott-Hubbard gap~\cite{sandilands2016} with a photon energy of $\hbar\omega$\,$\approx$\,$1.55$\,eV. 
The probe light has $2.42$\,eV photon energy. Under zero-field conditions, two contributions to the total optical polarization rotation $\theta_{\rm tot}$ can be distinguished: 

\begin{equation}
\theta_{\rm tot}= \theta_{\rm LD} + \theta_{\rm MLD}(\vec{L}^2).
\label{eq:totalrotation}
\end{equation}

\noindent The first term $\theta_{\rm LD}$, linear dichroism, originates from the monoclinic distortion of \RC, \cite{johnson2015} and will only show a negligible temperature dependence over the relevant temperature range.\cite{glamazda2017}
The second term $\theta_{\rm MLD}$, magnetic linear dichroism (MLD), \cite{smolenskiui1975birefringence,pisarev1991optical} is proportional to the square of the zigzag antiferromagnetic order parameter $\vec{L}=\vec{M_{\uparrow}}-\vec{M_{\downarrow}}$, where $\vec{M_{\uparrow}}$ and $\vec{M_{\downarrow}}$ give the sublattice magnetizations.
As such, the MLD rotation provides an optical probe of the zigzag spin order in \aRC.

\begin{figure}[h!]
 \center
\includegraphics[scale=1]{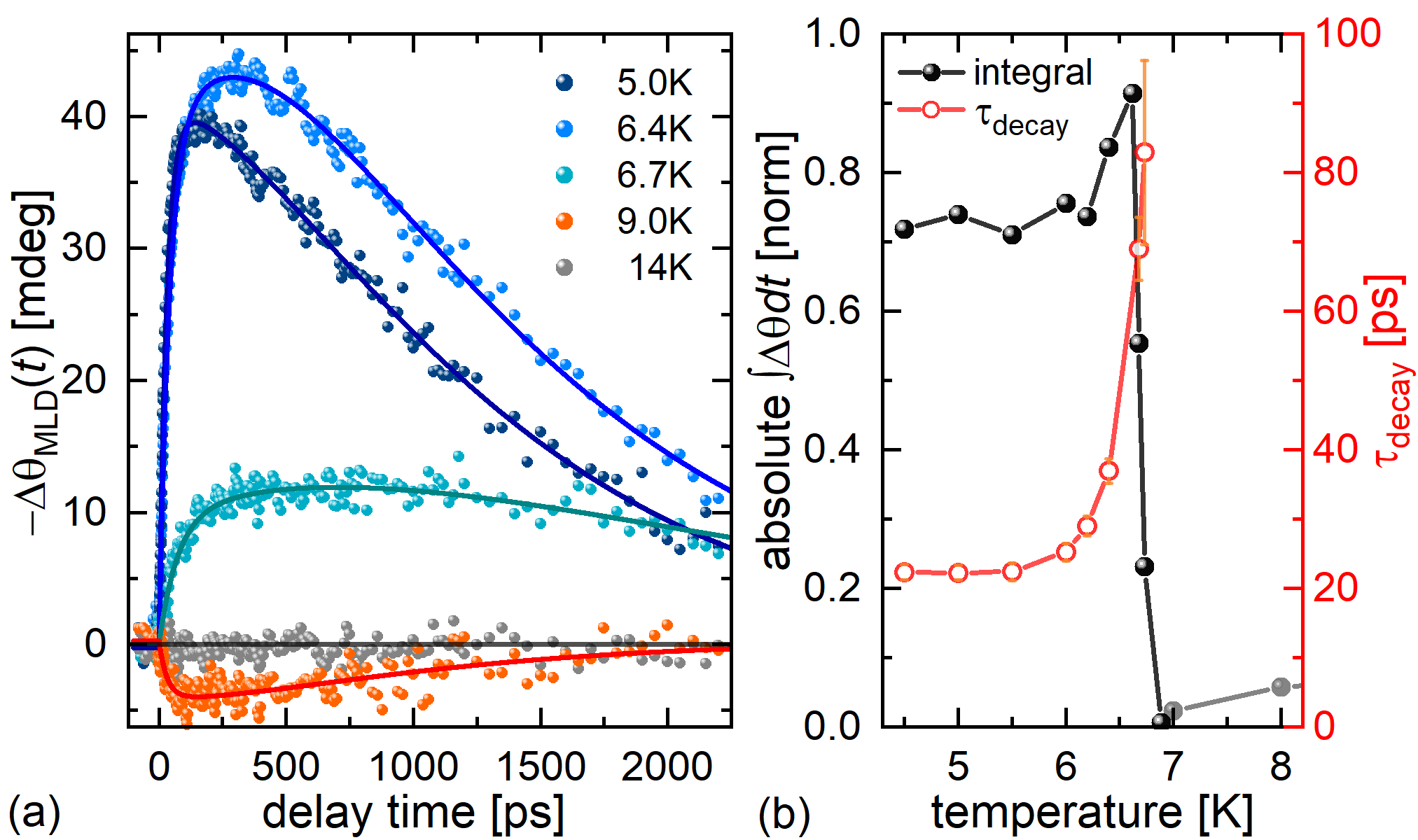}   
\caption{Temperature dependent transient polarization rotation and critical slowing down. a) Photoinduced change in polarization rotation $-\Delta\theta_{\rm MLD}(t)$ for various temperatures below and above $T_{\rm N}$\,$\approx$\,$7$\,K. The signal below $T_{\rm N}$ is dominated by the proper stacking phase. Above $T_{\rm N}$ a small signal with opposite rotational sign is observed, originating from the stacking-fault phase. b) Integrated change in rotation (black spheres) and $\tau_{\rm decay}$[ps] (red circles) as a function of temperature. A critical slowing down of the disordering is observed upon approaching the phase transition.}
\label{fig:tempdep}
\end{figure}

Figure \ref{fig:tempdep}a displays the photoinduced change in polarization rotation $-\Delta\theta_{\rm MLD}(t)$ for various bath temperatures. A low-excitation fluence $F$\,$\sim$\,$1.7$\,$\mu$J/cm$^{2}$ was used, corresponding to a photoexcitation density of $n_{\gamma}$\,$\approx$\,$0.8$\,$\cdot$\,$10^{17}$\,cm$^{-3}$ (Ref.\,\citenum{footnoteS1}). For temperatures below $T_{\rm N}$\,$\approx$\,$7$\,K an initial fast demagnetization on the 
tens of ps timescale is observed, after which the signal recovers on the ns-timescale. Above $T_{\rm N}$ a small amplitude response is observed with an opposite rotational sense, originating
from a fraction of unavoidable stacking-fault-phase contributions at the sample surface.\cite{sandilands2016,cao2016}
In Fig.\,\ref{fig:tempdep}b the integrated change in rotation $\Delta\theta_{\rm max}$ is plotted versus temperature. The integrated rotation change shows a pronounced increase, followed by a rapid reduction upon approaching $T_{\rm N}$\,$\approx$\,$7$\,K. This behavior is qualitatively rationalized by considering that the photoexcitation will have the largest transient effect where the derivative of the zigzag order parameter with respect to temperature is the largest~\cite{banerjee2017neutron}. Concomitantly, we observe a critical slowing down of the disordering upon approaching the phase transition~\cite{hohenberghalperin1977,zong2019}. This behavior is well captured by a $\tau_{\rm decay}$\,$\propto$\,$\vert 1-T/T_{\rm N}\vert^{-\nu z}$ power law with critical exponent $\nu z$\,$=$\,$-2.1$, compatible with the universality class of the 2D Ising model-A dynamics, applicable to \aRC~\cite{banerjee2017neutron,hohenberghalperin1977,tauber2014critical}.

Figure \,\ref{fig:densitydep}a shows the transient rotation traces $\theta_{\rm MLD}(t)$ for various initial photoexcitation densities $n_{\gamma}$ (sphere symbols). 
The photoexcitation dependence of the maximum MLD change, $\Delta\theta_{\rm MLD,max}$, is depicted in Fig.\,\ref{fig:densitydep}b.   
Qualitatively, two excitation regimes can be distinguished. For lower excitation densities ($n_{\gamma}$\,$<$\,$n_{\rm \gamma,crit}$\,$\approx$\,$3$\,$\cdot$\,$10^{17}$\,cm$^{-3}$), the spin system partially disorders, followed by a subsequent recovery. In this regime the disordering time slows down with increasing photoexcitation density. For the high excitation densities ($n_{\gamma}$\,$>$\,$n_{\rm\gamma, crit}$\,$\approx$\,$3$\,$\cdot$\,$10^{17}$\,cm$^{-3}$) a faster disordering time is observed and
the change $\Delta\theta_{\rm MLD,max}$ saturates (Fig.~\ref{fig:densitydep}b), implying that the photoexcited system resides in a $L$\,=\,$0$ state for multiple $100$s of ps.
Referring to the magnetic phase diagram (Fig.~\ref{fig:phasediagram} and Refs.~\citenum{johnson2015,kasahara2018majorana}), this means that for quench strengths above $n_{\rm \gamma,crit}$ the zigzag 
order can be fully suppressed, leaving the system in a spin disordered state. 
The disordering mechanism and the nature of the long-lived transient state is corroborated below.

\begin{figure}[h!]
 \center
\includegraphics[scale=1]{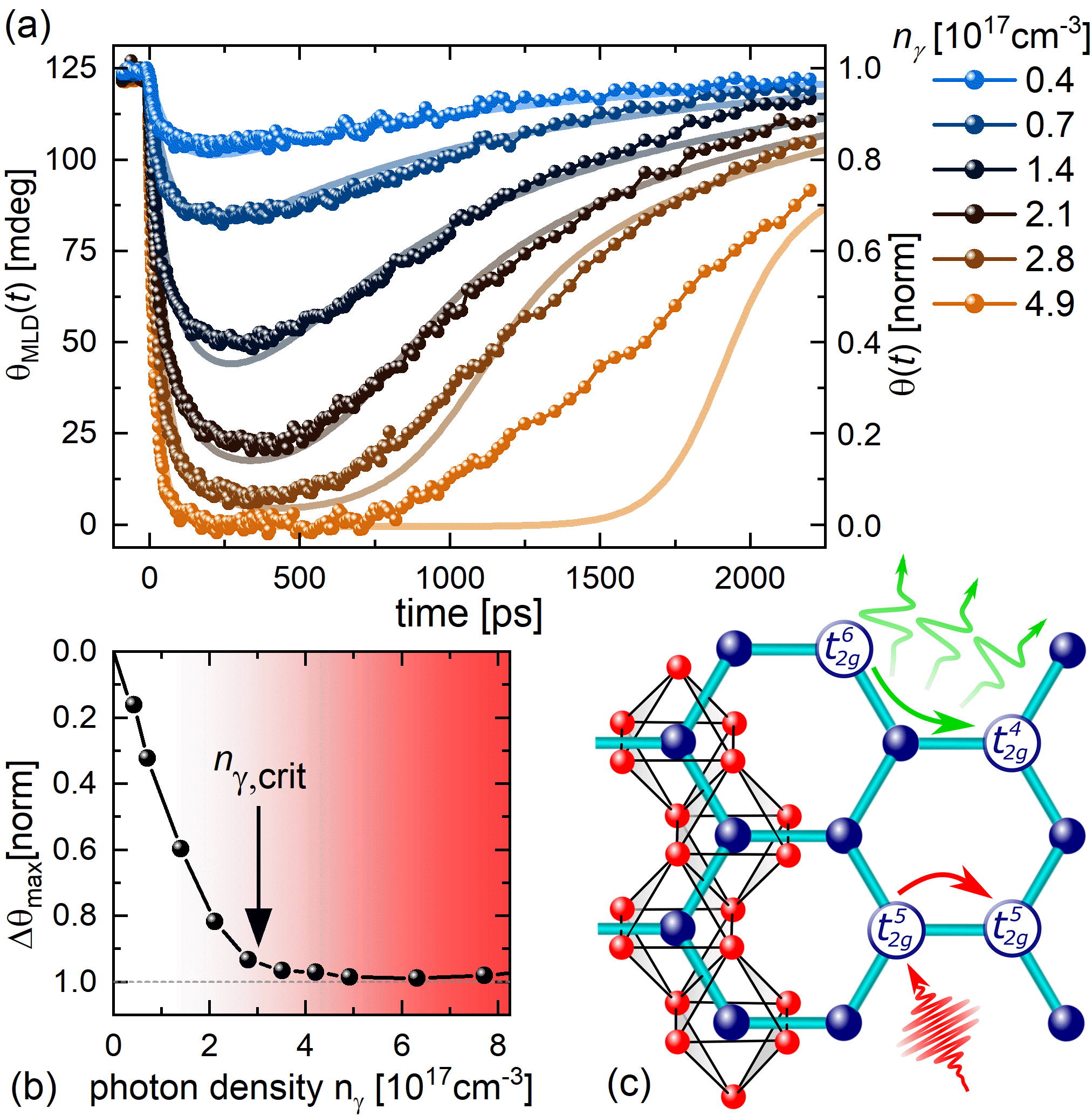}   
\caption{Nonequilibrium magnetic phase transition and holon-doublon pair recombination by multimagnon emission. a) Density-dependent $\theta_{\rm MLD}(t)$ for different excitation densities $n_{\gamma}$, as indicated with spheres. The modelled rotation $\theta(t)$ is indicated with thick lines. b) Maximum change in the magnetic linear dichroism (MLD) rotation $\Delta\theta_{\rm MLD}(t)$  as a function of photon density $n_{\gamma}$. Above the critical density $n_{\rm \gamma,crit}$\,$\approx$\,$3$\,$\cdot$\,$10^{17}$\,cm$^{-3}$ the maximum change in MLD-rotation saturates.
c) The honeycomb lattice, consisting of Ru-sites (dark-blue sites) and chloride ligand ions (red sites). The lower process
shows the photogeneration of a holon-doublon pair. The upper process shows the subsequent multimagnon emission by holon-doublon recombination.} 
\label{fig:densitydep}
\end{figure}

The inherently strong charge-spin coupling of Mott insulators leads to an efficient nonlinear demagnetization mechanism upon photoexcitation above the Mott-Hubbard gap.~\cite{lenarcic2013,afanasiev2019} 
In order to illustrate this mechanism, first consider the photoexcitation process corresponding to the lowest $t_{2g}^{5}$\,$t_{2g}^{5}$\,$\rightarrow$\,$t_{2g}^{4}$\,$t_{2g}^{6}$ hopping-type excitation across the Mott-Hubbard gap, as illustrated by the lower hopping process in Fig.~\ref{fig:densitydep}c. Within a quasiparticle picture, this intermediate excited state corresponds to a spinless \textit{holon} ($t_{2g}^{4}$) and \textit{doublon} ($t_{2g}^{6}$), by which effectively two magnetic moments are removed from the zigzag lattice. The mere \textit{creation} of these quasiparticles at the used low densities of $4$\,-\,$85$\,ppm photons/Ru$^{3+}$-site however does not suffice to explain the magnitude and timescale of the zigzag disordering.\cite{footnoteS1}  Instead, once created, the dominant decay mechanism of the holon-doublon pairs is  \textit{recombination} through multimagnon emission (upper hopping process Fig.\,\ref{fig:densitydep}c).\cite{lenarcic2013}
An order of magnitude estimate for the released amount of magnons per decayed $hd$-pair is provided by $\Delta_{\rm MH}/W$\,$\sim$\,$25$ (Refs.\,\citenum{lenarcic2013}), with $W$\,$\approx$\,$4.0$\,meV being
the bandwidth of the low-energy spin wave branch in the zigzag phase.\cite{banerjee2018excitations} As such, this quasiparticle recombination provides an efficient electronic demagnetization mechanism. 

In order to further delineate the excitation mechanism and resulting magnetization dynamics, we model the time-domain data within a dynamic Ginzburg-Landau (GL) model.\cite{hohenberghalperin1977,tauber2014critical} The holon-doublon density, representing the nonequilibrium dimension in Fig.\,\ref{fig:phasediagram}, comes in as a new dynamical variable here. We first consider the modified free energy for the antiferromagnetic order parameter $L$ and the holon-doublon-\textit{pair} density $n$: 
\begin{equation}
\mathcal{F}(n,L) =  \frac{a_1}{2}(n-n_{\rm c,eq}) L^2 + \frac{a_2}{4}L^4 + \tilde{\mathcal{F}}(n),
\end{equation}
\noindent with
\begin{equation}
\tilde{\mathcal{F}}(n)  =  a_3 n + \frac{a_4}{2} n^2 +\frac{a_5}{3}n^3 ,
\end{equation}

\noindent where $a_i$, $i=1, \dots, 5$ are phenomenological parameters.
The terms with even powers in the zigzag order parameter $L$ are the standard symmetry-allowed terms in the Landau free energy expansion for an antiferromagnet.\cite{tauber2014critical,khomskii2010basic} Notice that odd powers of $L$ are ruled out by the inversion symmetry of \RC.\cite{johnson2015} 
The initial value of the holon-doublon pair density $n(0)$, or quench strength, is taken proportional to the experimental photoexcitation densities $n_{\gamma}$, i.e., $n(0)$\,$\propto$\,$n_{\gamma}$, where each photon creates one $hd$-pair. 
The first term in $\mathcal{F}(n,L)$, coupling the $hd$-pair density $n$ to the order parameter $L$, leads to a destabilization of the magnetic order for a sufficiently strong excitation of $hd$-pairs, thus reproducing the process of annihilation of $hd$-pairs into magnons.\cite{lenarcic2013,footnoteS2}
The parameter $n_{\rm c,eq}$ is introduced as the critical $hd$-pair density at equilibrium. 
The functional $\tilde{\mathcal{F}}(n)$, independent of the order parameter $L$, describes the excess energy of the $hd$-density and its relaxation in the absence of magnetization. It therefore accounts for decay mechanisms other than the nonradiative multimagnon emission discussed above, such as nonradiative phonon emission, spontaneous decay under radiative emission,\cite{mitrano2014}, and possible $hd$-pair diffusion out of the probe volume.
The form of $\tilde{\mathcal{F}}(n)$ is chosen as a third-order polynomial, although its exact form is not crucial for the analysis.

The time evolution of the $hd$-pair density $n$ and magnetic order parameter $L$ is described by the coupled equations of motion:
\begin{equation}
\frac{d L}{dt}=-\frac{\delta \mathcal{F}}{\delta L}, \qquad \frac{d n}{dt}=-\frac{\delta \mathcal{F}}{\delta n}.
\label{eq:Onsager}
\end{equation}
In order to relate Eqs.\,\eqref{eq:Onsager} to the experimentally measured rotation $\theta_{\rm MLD}$, we rewrite the equations in terms of the polarization rotation $\theta$\,$=$\,$L^2/2$, to finally obtain:
\begin{subequations}
\label{eq:NEQ-Onsager}
\begin{align}
\frac{d \theta}{d t} &=-2a_1 (n-n_{\rm c,eq})\theta -4a_2\theta^2, \\
\frac{d n}{d t} & = -a_1\theta -\frac{\delta}{\delta n}\tilde{\mathcal{F}}(n)
\end{align}
\end{subequations}
By using Eq. \eqref{eq:NEQ-Onsager}a, the trajectories $n(t),\theta(t)$ can be modelled for different initial quench strengths $n(0)$, taken proportional to the experimental $n_{\gamma}$ densities. The curves for $\theta(t)$ are superimposed on the experimental $\theta_{\rm MLD}(t)$ in Fig.\,\ref{fig:densitydep}a. The model captures the dependence of the demagnetization time on the excitation density and the position of $t_{\rm max}$, i.e., the time at which $\Delta\theta_{\rm max}$ is reached. For the higher excitation densities the magnetic order vanishes, reproducing the long-lived transient $L$\,$=$\,$0$ state. The inclusion of the $n^3$-term in $\tilde{\mathcal{F}}(n)$ ensures that the GL-description does not overestimate the lifetime of the $L$\,$=$\,$0$ state.\cite{footnoteS3} The density-dependent rotation transients are well captured considering the minimal amount of parameters needed in the 
nonequilibrium GL-description.

The $hd$-pair density-dependent free-energy landscape $\mathcal{F}(n,L)$ is shown in Fig.~\ref{fig:freenenergy}. For low densities, the free energy retains its double-well profile, whereas for higher densities a single well forms.\cite{zong2019}
Representative trajectories $n(t),L(t)$ for different excitation densities are drawn into the free-energy landscape, with colors corresponding to the conceptual trajectories of Fig.~\ref{fig:phasediagram}.  
The quench $n(0)$ brings the system into a high-energy state, after which $n$ and $L$ relax along the minimal energy trajectory. For a small quench (red trajectory)
the zigzag order parameter $L(t)$ stays finite and eventually recovers. For the intermediate densities (magenta trajectory)
the $n(t),L(t)$ coordinates approach the $L$\,$=$\,$0$ line. For the higher excitation densities (light blue trajectory)
the $hd$-pairs have sufficient excess energy to let $n(t),L(t)$ follow a trajectory along the $L$\,$=$\,$0$ line, i.e., full spin disordering is reached. We emphasize that, for strong quenches, the excitation density $n(t)$ still varies in time, even though $L(t)$ takes the quasistationary value $L(t)$\,$=$\,$0$.
A sufficiently strong photoexcitation quench thus provides a mechanism to dynamically stabilize a \textit{nonequilibrium quasistationary} spin disorded state in \aRC.

\begin{figure}[h!]
 \center
\includegraphics[width=3.375in]{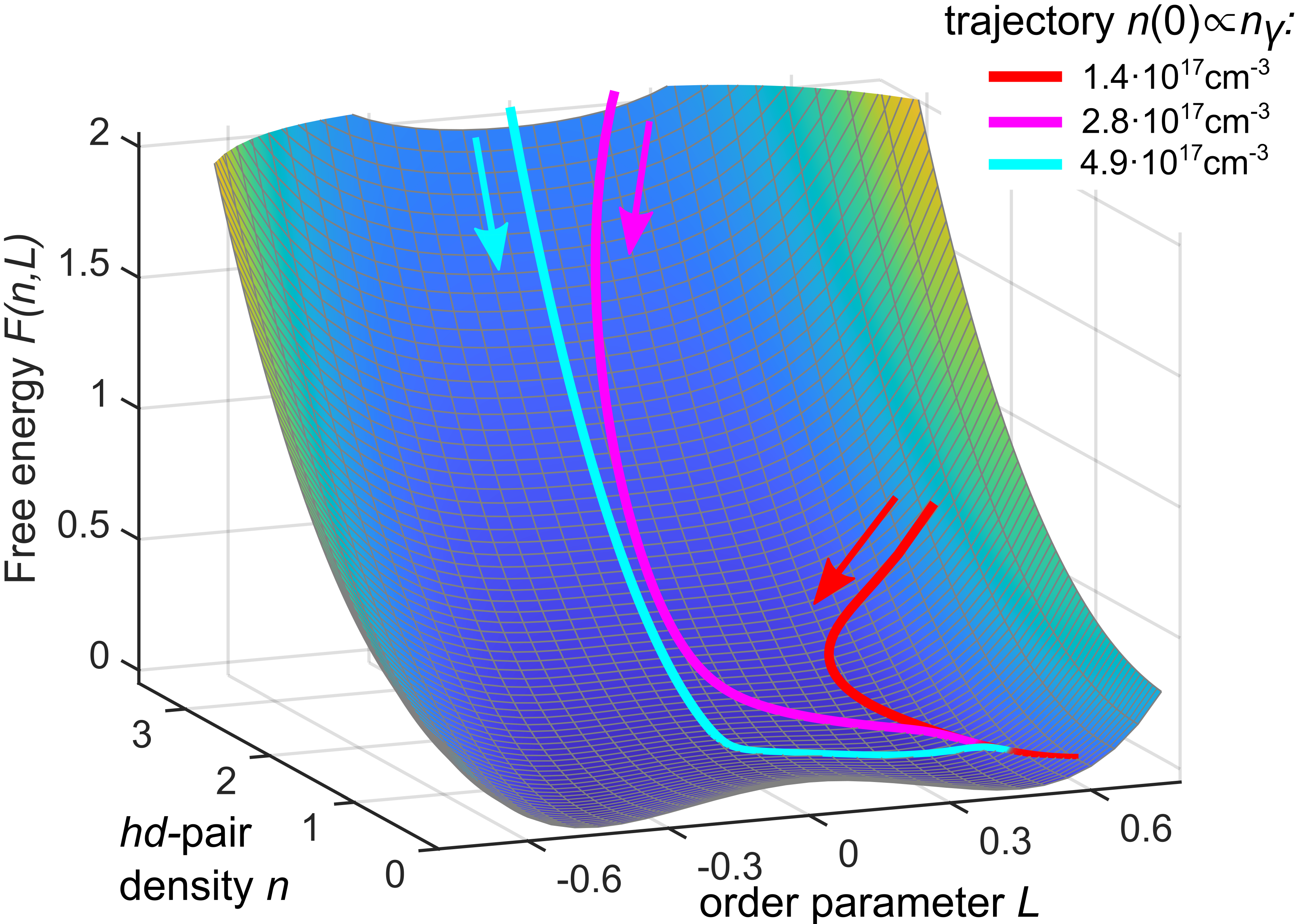} 
\caption{Free energy landscape and nonequilibrium quasistationary spin disordered state. Free energy landscape $F(n,L)$ as a function of the zigzag order parameter $L$ and $hd$-pair density $n$ (cf. the phase diagram in Fig.\,\ref{fig:phasediagram}). The initial quench $n(0)$ brings the system to a high energy state, after which the system relaxes. For small quenches (red trajectory), the order parameter $L$ stays finite under the relaxation of the density $n$. For intermediate quenches (magenta trajectory) the system approaches the $L$\,=$0$ line. For strong quenches (light blue trajectory, highest energies not shown) the system relaxes along the $L$\,=$0$ line, implying that the system is described as a nonequilibrium quasistationary spin disordered state.}
\label{fig:freenenergy}
\end{figure}

The maximum lifetime of the nonequilibrium quasistationary spin disordered state is dictated by the recombination rate of the $hd$-pairs. The time evolution of $n(t)$ obtained from the nonequilibrium GL model provides us with an estimate of a few nanoseconds for the recombination timescale of the $hd$-pairs.\cite{footnoteS3}
This timescale
is expected to grow
exponentially with the number of magnons needed to traverse the Mott gap, i.e., $\tau$\,$\sim$\,$e^{\Delta_{\rm MH}/W}$. \cite{lenarcic2013} 
Considering the weak exchange-interaction scale $W$ in \aRC, one may indeed expect significantly longer recombination times compared to materials with an order of magnitude stronger exchange, such as Nd$_2$CuO$_4$ and Sr$_2$IrO$_4$, where $hd$-pair lifetimes on the order of $0.1$\,ps have been reported.\cite{afanasiev2019,okamoto2011} A large ratio between the Mott gap and the exchange interaction energy thus is the key element to ensure a long lifetime of the nonequilibrium quasistationary state.

The microscopic nature of the transient long-lived spin disordered state currently remains elusive. The used low excitation densities by far do not provide sufficient energy to drive the material into a conventional paramagnetic state, nor to change the dominant interactions in the system. Considering the phase diagram, it therefore seems plausible that the system is driven into a transient proximate spin liquid phase, reminiscent of the thermodynamic state just above $T_{\rm N}$. Energy-resolved ultrafast techniques may provide more insight into the microscopic properties of the induced phase. \cite{nembrini2016,sandilands2015raman,halasz2016} Furthermore, exact diagonalization \cite{okamoto2013} and nonequilibrium dynamical mean-field theory \cite{aoki2014} methods may elucidate the role of $hd$-excitations in the Kitaev-Heisenberg model and the resulting phase diagram.

\section*{Conclusions}

We have unveiled a pulsed light excitation driven mechanism allowing to trap a Kitaev-Heisenberg magnet into a quasistationary spin disordered state. Photoexcitation above the Mott-gap generates a transient density of holon-doublon quasiparticle pairs. The subsequent recombination of these quasiparticles through efficient multimagnon emission provides a way to dynamically destabilize the competing zigzag ordered ground state and thereby keeps the system in an out-of-equilibrium spin disordered state, up until the transient electronic quasiparticle density gets depleted. Our work provides insight into the coupling between electronic and magnetic degrees of freedom in \aRC, and suggest a new way to reach a proximate spin liquid phase in Kitaev- Heisenberg magnets. 

\section*{Materials and methods}
\paragraph*{Sample growth and characterization} High-quality \aRC crystals were prepared by vacuum sublimation.\cite{sahasrabudhe2019high} Different samples of the batch were characterized by SQUID magnetometry, showing a sharp phase transition at $T_{\rm N}$\,$\approx$\,$7$K. This bulk technique can only provide a first indication of sample quality for an optics study. Cleaving or polishing of \RC samples introduces strain, which leads to stacking faults. For the optics study, we therefore refrained from any sample treatment, and used an as-grown \RC sample with a shiny $\sim$\,$1.5$\,$\times$\,$1.5$\,mm$^2$ surface area. The temperature dependence shown in Fig.\,\ref{fig:tempdep}b shows a clear phase transition at $T_{\rm N}$\,$\approx$\,$7$K. 

\paragraph*{Time-resolved magneto-optical experiment}
The \aRC sample is mounted in a bath cryostat. The time-resolved magneto-optical experiment was performed using $800$\,nm pump pulses with a temporal with of $40$\,fs, and probe pulses of $512$\,nm with a temporal width of $250$\,fs. The pump and probe beam were focused down to a radius of $r_{\rm pump}$\,$\approx$\,$39$\,$\mu$m and $r_{\rm probe}$\,$\approx$\,$25$\,$\mu$m, respectively. The repetition rate of the amplified laser system was set to $f$\,$=$\,$30$\,kHz in order to ensure that the system can relax back to the ground state between consecutive pulses. The change in polarization rotation of the reflected probe pulse is measured via a standard polarization bridge scheme. 
The optical conductivity reported in Ref.\,\citenum{sandilands2016} and the structural properties reported in Ref.\,\citenum{johnson2015} allows us to calculate the photoexcitation densities, as outlined in more detail in the Supplementary Material.\cite{footnoteS1}

\section*{Acknowledgements}
The authors thank A.~Rosch (Cologne, DE) and Z.~Lenar\v{c}i\v{c} (Berkeley, USA) for fruitful discussions.
This project was partially financed by the Deutsche
Forschungsgemeinschaft (DFG) through Project No.
277146847 - Collaborative Research Center 1238: Control
and Dynamics of Quantum Materials (Subprojects No. B05
and No. C04) and through project INST 216/783-1 FUGG.
S.D. acknowledges support by the European
Research Council (ERC) under the Horizon 2020 research and innovation program, Grant Agreement No. 647434 (DOQS).

\end{document}